\documentclass[a4paper,fleqn,usenatbib]{mn2e}

\usepackage{newtxtext,newtxmath}

\usepackage[T1]{fontenc}
\usepackage{ae,aecompl}
\usepackage{footnote}

\usepackage{graphicx}	
\usepackage{amsmath}	
\usepackage{amssymb}	
\usepackage[caption=false]{subfig}
\usepackage{threeparttable}
\usepackage{multirow}

\def \st{\ifmmode{  {\mathrm{st}}}\else{${  {\mathrm{st}}}$}\fi}
\def \nd{\ifmmode{  {\mathrm{nd}}}\else{${  {\mathrm{nd}}}$}\fi}
\def \rd{\ifmmode{  {\mathrm{rd}}}\else{${  {\mathrm{rd}}}$}\fi}
\def \th{\ifmmode{  {\mathrm{th}}}\else{${  {\mathrm{th}}}$}\fi}

\title[The Central Star of Planetary Nebula PHR~1315-6555 and its host Galactic Open Cluster AL~1]
{\centering The Central Star of Planetary Nebula PHR~1315-6555 and its host Galactic Open Cluster AL~1}

\author[V. Fragkou et al.]{V. Fragkou,$^{1,2}$\thanks{vfrag@hku.hk} Q. A. Parker,$^{1,2}$\thanks{quentinp@hku.hk}
A. A. Zijlstra,$^{3}$ R. Shaw,$^{4}$ and  F. Lykou$^{1,2}$
\\
 $^{1}$Department of Physics, The University of Hong Kong, Hong Kong SAR, China\\
$^{2}$Laboratory for Space Research, The University of Hong Kong, Hong Kong SAR, China\\ 
 $^{3}$The University of Manchester, Manchester, UK\\
$^{4}$Space Telescope Science Institute, Maryland, USA\\
}

\date{Accepted XXX. Received YYY; in original form ZZZ}

\pubyear{2018}

\begin{document}
\label{firstpage}
\pagerange{\pageref{firstpage}--\pageref{lastpage}}
\maketitle

\begin{abstract}

PHR~1315-6555 is a rare case of a Galactic Planetary Nebula that is a proven member of the Open Cluster AL~1. This allows its distance to 
be defined with precision and thus the accurate measurement of its physical characteristics along with the parameters of its Central Star (CS). In this work we 
use HST to detect this unique CS and constrain the cluster's physical parameters. Our results suggest that the 
cluster rests at a distance of $\sim$12 kpc, is highly reddened, and has 
an age of around 0.66~Gyrs and a turn-off mass of $\sim$2.2 $M_\odot$. Our deep Colour Magnitude Diagram (CMD) suggests that the metallicity of the cluster is subsolar (Z=0.006). Our 
photometric measurements indicate that the PN's core is a faint blue star close to the nebular apparent centre, with an observed dereddened visual 
VEGA magnitude of 21.82 $\pm$ 0.60. A significant contribution from any possible binary companion is unlikely but possible. Our results show that the CS has an 
effective Zanstra temperature of around 113 kK and a mass of 0.58$M_\odot$ providing a unique additional point to the fundamental White Dwarf 
Initial-to-Final-Mass Relation. 

\end{abstract}



\section{Introduction}

At the end of their lives, low-to-intermediate mass stars ($\sim$1 to $\sim$8 $M_\odot$) pass through the AGB phase, at which point they lose most of their mass and if their remnant stellar cores reach temperatures high enough to ionise the ejected material, a Planetary Nebula (PN) is formed around them. 
These Central Stars of Planetary Nebulae (CSPNe) evolve at constant luminosities towards higher effective temperatures. After their fuel is exhausted they will 
eventually cool along the White Dwarf (WD) cooling track. Their evolution depends on the thermal pulse cycle phase during which the stars left the AGB phase 
(Schonberner 1983; Vassiliadis \& Wood 1994; Blocker 1995).

As the AGB mass-loss is mainly dust-driven (Wood 1979; Bowen 1988) and the formation of the dust depends strongly on metallicity, high metallicity AGB stars 
lose a larger amount of matter compared to stars of lower metallicities (Willson 2000) and consequently, result in less massive CSPNe (see Villaver et al. 2003; 
Villaver, Stanghellini \& Shaw 2004). Furthermore, in the case of mass-loss under lack of dust, low metallicity AGB stars are physically smaller for a given 
luminosity and mass (Willson, Bowen \& Struck 1996; Willson 2000) and as a result their mass-loss rates are reduced (see Villaver et al. 2004). This dependency 
of mass-loss rates has as a consequence that in low metallicity environments and for a given initial stellar population more Main Sequence (MS) stars reach the 
Chandrasekhar limit and produce Type II Supernovae (see Villaver et al. 2004). 

PNe are visible only for a short  time period ($\sim$10,000-30,000 yrs, Badenes, Maoz, \& Ciardullo 2015) before their ionised material dissipates. Their 
evolutionary timescales depend on the progenitor masses of their CSPNe (e.g. Villaver, Manchado \& Garcia-Segura 2002). CSPNe masses are crucial in 
understanding post-AGB evolution as they provide additional data for the widely used WD initial-to-final mass relation (IFMR; e.g. Ferrario et al. 2005; Dobbie et 
al. 2009; Kalirai et al. 2008) and information about the dredge-up efficiency during the AGB phase (see Parker et al. 2011; Moreno-Ibanez et al. 2016). A reliable 
IFMR is a powerful tool in using WD luminosity functions for estimating ages of the Galactic disk and of open clusters using field WDs and cluster WD populations 
respectively. They can help us trace the enhancement of both nitrogen and carbon in entire galaxies (see Parker et al. 2011). The IFMR has some correlation with 
metallicity (Weidemann 1987; Vassiliadis \& Wood 1994; Marigo \& Girardi 2007; Miller Bertolami 2016) as the upper mass limits for WD production is lower in 
metal poor environments (see Villaver et al. 2003). Moreover, the mass-loss and the convection mechanisms constrain the upper limits for the 
initial mass of stars that will evolve into WDs (Blocker 1995; Herwig 2000) but both these processes are not well understood. Thus any precise measurements of 
CSPN and WD masses are of major importance (see e.g. Villaver, Stanghellini \& Shaw 2003).

The initial mass of a CSPN can also drive to the chemistry of the resulting PN as higher mass progenitors tend to form PNe with enhanced N, which are 
usually of axisymmetric or bipolar morphologies (Type~I PNe, Peimbert 1978; Peimbert \& Torres-Peimbert 1983). This is corroborated by the fact that Galactic 
bipolar PNe tend to be found at low galactic latitudes (Corradi \& Schwarz 1995; Manchado et al. 2000; Stanghellini et al. 2002; Parker et al. 2006). From 
hydrodynamic modelling of such massive progenitor stars (e.g. Villaver et al. 2002; Perinotto et al. 2004) we expect Type~I PNe to be optically thick (e.g. Kaler \& 
Jacoby 1989) for the majority of their lifespans. In some cases they may never turn into optically thin nebulae, though bipolar PNe may be optically thin in their 
lobes and thick in the torus (see Moreno-Ibanez et al. 2016). As a result PNe morphologies can provide us hints regarding their physical properties (e.g. 
Stanghellini, Shaw \& Villaver 2016). 

Despite their importance for stellar evolution, CSPNe studies are difficult since their inherent luminosities are low and they are often too faint to be easily detected 
compared to the surrounding nebula (Shaw \& Kaler 1985). Measurements of their masses and other characteristics require a precise (accuracy better than 10\%, 
Shaw 2006) determination of their distances (see Villaver, Stanghellini \& Shaw 2007). Although the accurately known distances of external galaxies allow the 
study of their PNe and in some cases their CSPNe (e.g. Villaver et al. 2003; Villaver et al. 2004), such a task is extremely difficult for PNe in our Galaxy since only a few Galactic PNe distances are determined with sufficient precision (see Moreno-Ibanez et al. 2016). Precise parallax measurements currently exist only for a very small fraction (sigma/parallax=$\sigma$/$\pi$  $\sim$5\%, Benedict et al. 2009) of close by Galactic PNe, and large statistical uncertainties ($\sim$20 -  30\%, Stanghellini, Shaw \& Villaver 2008; Giammanco et al. 2011; Frew, Parker \& Bojicic 2016) affect the distance estimates of bulk PNe (see Majaess et al. 2014), though this situation is likely to change in the near future, at least for close by PNe, with the complete data release from the GAIA mission (Gaia Collaboration et al. 2016).

PNe that are members of Galactic stellar clusters present the advantage of allowing an accurate (< 10\%) determination of their distances from cluster Colour-
Magnitude Diagrams (CMDs) and precise measurements of their physical properties such as their ages, physical dimensions, chemical composition (as this can 
be independently considered from the host cluster's metallicity), effective temperatures and masses of their progenitor stars from fits to cluster isochrones (Parker 
et al. 2011; Turner et al. 2011; Moni Bidin et al. 2014). Furthermore, photometric measurements of their CSPNe can constrain their intrinsic luminosity and mass 
and thus, these objects can be used as additional points for the IFMR (Parker et al. 2011). 

Unfortunately, evolved stars going through the PN phase in open clusters are very hard to find. This is because open clusters usually survive for less than 1~Gyr (Bonatto 
\& Bica 2011) so by the time the dominant low mass stars enter the PNe stage the cluster will have completely dissipated.   Clusters with longer lifespans are 
generally more massive and so can host more numbers of massive stars. However, the more massive of these do not evolve as PNe (e.g. Majaess, Turner \& 
Lane 2007). The lifetimes of PNe from progenitor stars of a few solar masses in young clusters are short, only around $10^3$- $10^4$ yrs (see Majaess et al. 
2014) making the chance of their detection unlikely. Hence, any examples uncovered are rare jewels for scientific exploitation. Until today, only five PNe have been 
found to be physically associated with stellar clusters in our Galaxy. However, four are found in extremely long-lived Globular Clusters (Pease et al. 1928; Gillet et 
al. 1989; Jacoby et al. 1997). So far only one (PHR~1315-6555) has been proven to be a member of  an intermediate age Open Cluster (OC hereafter, Parker et 
al. 2011) with a turn-off mass of $\sim$2.2$M_\odot$. 

PHR~1315-6555 is a faint bipolar, likely Type~I, PN discovered through the AAO/UKST SuperCOSMOS H$\alpha$ survey (Parker et al. 2005), whose radial-
velocity, interstellar extinction and statistical distance measurements show that it is a member of the distant Galactic OC Andrews- Lindsay 1 (AL~1). It has an 
apparent angular diameter of 80~arcseconds (Parker et al. 2011; Majaess et al. 2014). 

The cluster isochrone derived distance for this PN of 10 kpc has been estimated to be the most accurate currently determined for a PN in our Galaxy (sigma/
distance=$\sigma$/d = 4\%, Majaess et al. 2014) and although it puts it beyond the reach of GAIA  (Gaia Collaboration et al. 2016), this allows the direct study if 
its CSPN. 

In this work we present our deep HST F555W and F814W photometry of the cluster, the PN and its CSPN. This has allowed us to create an improved CMD of the 
cluster, explore the nebular microstructure and determine the physical properties of its CSPN, a unique and rare addition for the WD IFMR.  In Section 2, we 
investigate the physical properties of the OC AL~1, while in Section 3, we examine the PN and its CSPN. Finally, in Section 4 we discuss our results and in 
Section 5 we present our conclusions.  

\section{The Open Cluster Andrews - Lindsay 1 (AL~1)}

AL~1 is a distant and faint compact OC that lies close to the solar circle (see Carraro, Vallenari \& Ortolani 1995), first detected by Andrews \& Lindsay (1967) and 
van den Bergh \& Hagen (1975). Its ESO Schmidt plates designation is ESO 96-SC04 (Lauberts 1982) and various authors have explored its properties (Phelps, 
Janes \& Montgomery 1994; Janes \& Phelps 1994; Carraro at al. 1995; Carraro \& Munari 2004; Frinchaboy et al. 2004a,b; Carraro, Janes \& Eastman 2005; 
Majaess et al. 2014). 

Janes \& Phelps (1994) found a 7.5 kpc distance to the cluster using the mean luminosity of its red giant clump but considering its sparseness this is probably an 
underestimation of the cluster's true distance (Carraro et al. 1995). Using the photometric data from Phelps et al. (1994) for the calibration of their frames, Carraro 
et al. (1995) found that the best fit of their B and V observed magnitudes of 2059 cluster stars on the Padova isochrone scales (Girardi et al. 2000) predicts a 
cluster age of 0.7~Gyrs and a reddening E(B-V) $\sim$ 0.75 mag. This agrees within the errors with the value obtained by Neckel \& Klare (1980) for the visual 
absorption at the direction of the cluster ($A_{\rm V}$ = 1.7 - 1.9 mag). Although their dereddened Turn Off (TO) colour $(B-V)_0$ indicates that the cluster's 
metallicity (Z) is probably slightly lower than solar (also supported from abundance measurements [Fe/H] found to be -0.51 $\pm$ 0.3 by Frinchaboy et al. 
2004a), the isochrones of Z $=$ 0.008 do not fit their data. Thus, assuming solar metallicity, they derived a distance to the cluster of around 11.8 kpc. This leads 
to cluster Galactic coordinates of X $=$ -9.6 kpc, Y $=$ 1.7 kpc and Z $=$ -0.7 kpc (Carraro et al. 1995). 

Carraro \& Munari (2004) collected BVI photometric data for 890 cluster stars  and also assuming a solar metallicity found a cluster age of 0.8~Gyrs, 
distance of 12 $\pm $1 kpc and a reddening of 0.7 $\pm$ 0.2 mag, in a close agreement with the results of Carraro et al. (1995). Their results indicate that the 
progenitor mass of PHR1315-6555 is around 2.5$M_\odot$ (e.g. Girardi et al. 2000) slightly depending on Z, while Majaess et al. (2014) has estimated a TO 
mass for the cluster of 2.3$M_\odot$.

The cluster physical parameters derived by the different authors are summarized for convenience in Table 1. Their large spread reflects the difficulty of studying a 
faint and distant cluster suffering from large contamination from field stars (see Majaess et al. 2014). 
The more disparate results  by Carraro et al. (2005) were undertaken 
under poor weather conditions. In the following we calculate afresh the parameters of this faint cluster using our deep HST F555W F814W photometry. This has 
enabled an improved CMD that extends 4 to 5 magnitudes fainter than any previously obtained for this particular cluster.

\begin{table*}  
\begin{threeparttable}
\center
\caption[]{The physical parameters of AL~1 as determined from this work and from previous authors. }  
\label{table1}

\begin{tabular}{lccccll}  
\noalign{\smallskip}  

\hline
\hline

Distance  (kpc) & E(B-V) & Age (Gyrs) & Reference \\

\hline

12 $\pm$ 0.5 & 0.83 $\pm$ 0.05 & 0.66 $\pm$ 0.10 & This work  \\
10 $\pm$ 0.4 & 0.72 & 0.794 $\pm$ 0.106 & Majaess et al. (2014)  \\
16.95 & 0.34 $\pm$ 0.05 & 0.8 $\pm$ 0.2 & Carraro et al. (2005)  \\
12 $\pm$ 1 & 0.7 $\pm$ 0.2 & 0.8 & Carraro \& Munari (2004)  \\
9.35 & -  & 0.67 &   Frinchaboy et al. (2004b)\\
11.8 & 0.75 & 0.7 & Carraro et al. (1995)  \\
7.57 & 0.72 & - & Janes \& Phelps (1994)  \\

\hline

\end{tabular}
\end{threeparttable}
\end{table*}  

\subsection{Observations and data reduction}

Under Program ID: 12518 (06 March 2012), we obtained both long and short time exposures in each of the HST WFC3 (Wide-Field-Camera 3, Kimble et al 2008) 
F555W and F814W filters, centred on the cluster's apparent centre (see Fig. 1). This was in order to cover the full dynamic range of the cluster's stars with S/N $
\geqslant$ 30 from above the red giant clump (V $\sim$ 16) to the faint end of the cluster's luminosity function (V $\sim$ 26). These two filters measure the V and 
I continuum respectively and have passbands very similar to those of the Johnson-Cousins system. As the WFC3 field of view (162 $\times$ 162~arcseconds) is nearly identical to the size of the cluster these observations allow the construction of a CMD that samples effectively the entire cluster. The complete observing log can be seen in Table 2.

\begin{table}  
\begin{threeparttable}
\center
\caption[]{Imaging log. for HST program ID: 12518 }  
\label{table2}

\begin{tabular}{lccccl}  
\noalign{\smallskip}  

\hline
\hline
\multicolumn{5}{c}{HST Observing log. for the field on 06/03/2012} \\  
\hline  

 \: \: $\alpha$ & $\delta$ & $t_\text{exp}$ & Filter & \: \:\: $\lambda_{\rm c}$ \\
(h m s) & (\degr\ \arcmin\ \arcsec) & (sec) &  & \:\:\: (\AA) \\  
\hline

13:15:18.90 & -65:55:01.00 & 1000 & F502N & 5009.60  \\
13:15:16.00 & -65:55:16.00 &  1020 & F555W & 5305.95  \\
\ldots & \ldots &  1020 & F5814W & 8048.10  \\
\ldots & \ldots & 1020 & F200LP & 4939.20  \\
\ldots & \ldots  & 1100 & F350LP & 5871.50  \\
\ldots & \ldots  & 36 & F555W & 5305.50  \\
\ldots & \ldots  & 40 & F814W & 8043.70  \\

\hline

\end{tabular}

\end{threeparttable}
\end{table}  

The pixel size of the UVIS channel of WFC3 is 0.04~arcseconds per pixel (Dressel 2012) and observations have been made with a gain of 1.5 $e^-$/ADU. The 
images have been processed using the standard WFC3 calibration pipeline (CALWFC3 version 3.4.1, 10 Apr 2017; 
the full calibration process described by Rajan et al. 2011).

For the analysis of our data and measuring the magnitudes of the cluster's stars, the IRAF/DAOPHOT package was used (Stetson 1987; Davis 1994). DAOPHOT 
requires an initial estimate of the FWHM and thus we obtained a FWHM estimate of around 4 pixels for the stellar profiles of our long exposures and around 2.8 for 
our short exposures. We identified the stars in our images, measured their instrumental magnitudes and calculated and fitted a Point Spread Function (PSF). We 
considered the fraction of the stellar PSF that falls outside the measured stellar apertures to perform aperture correction. We finally transformed the measured 
values to the Space Telescope magnitude system (STMAG), which is based on a spectrum with constant flux per unit wavelength
following Dressel (2017). As the WFC3/UVIS channel presents a variable PSF (Sabbi \& Bellini 2013), around 20-30 bright and relatively isolated stars spread around each image, were used for its calculation. The detailed processing steps of IRAF/DAOPHOT are described by Artusi et al. (2016). The zero-point calibration (= -21.10 as provided by HST) for the transformation to ST magnitudes is described by Koornneef et al. (1986) and Horne (1988).

Our data have been checked for consistency by comparing our measured F555W magnitudes with the visual magnitudes measured for the same stars by 
Majaess et al. (2014). In Fig. 2 we can see that our magnitudes agree with those previously measured for the stars of the cluster taking into account the difference 
in the magnitude system used. The apparent scatter can be explained by the fact that lower resolution ground based observational data are more affected by 
blending in a crowded field such as the one studied here. Finally, we excluded saturated stars in our images and removed duplications for magnitudes obtained from 
both our short and long exposures in both filters. Our measured magnitudes were then transformed into the VEGA magnitude system VEGAMAG using the corresponding zero points provided by the HST WFC3 handbook (Rajan et al. 2011).

\begin{figure*}
\centering
\includegraphics[scale=0.2]{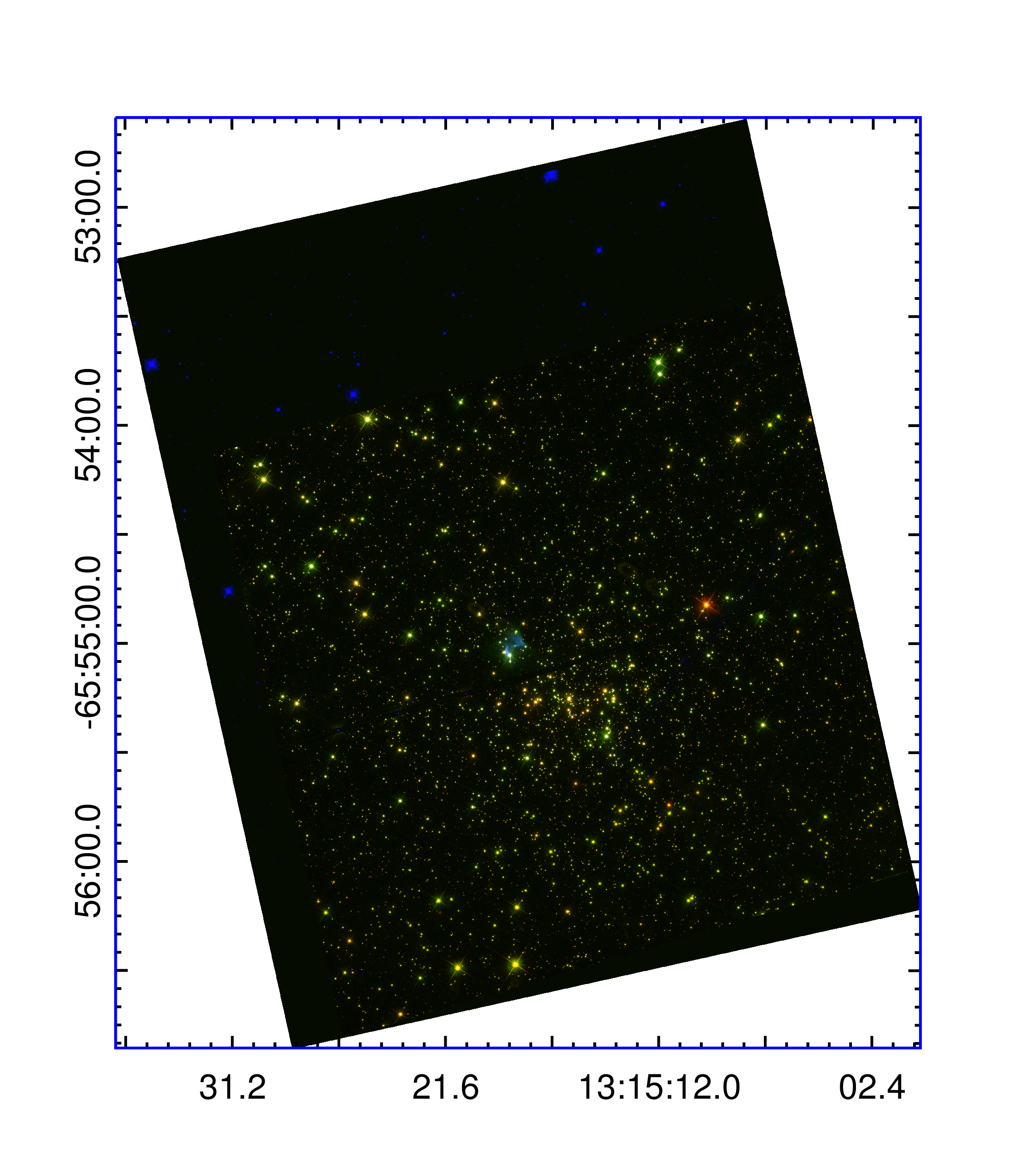}
\caption{A colour composite image of our HST field of view of OC AL~1. The F502N narrow filter image is blue, the F555W long exposure image in green and the F814W long exposure image is red. Colours have been adjusted in each RGB channel to try to represent the natural star colours.}
\label{fig1}
\end{figure*}

\begin{figure*}
\centering
\begin{minipage}[b]{.4\textwidth}
\includegraphics[width=\textwidth]{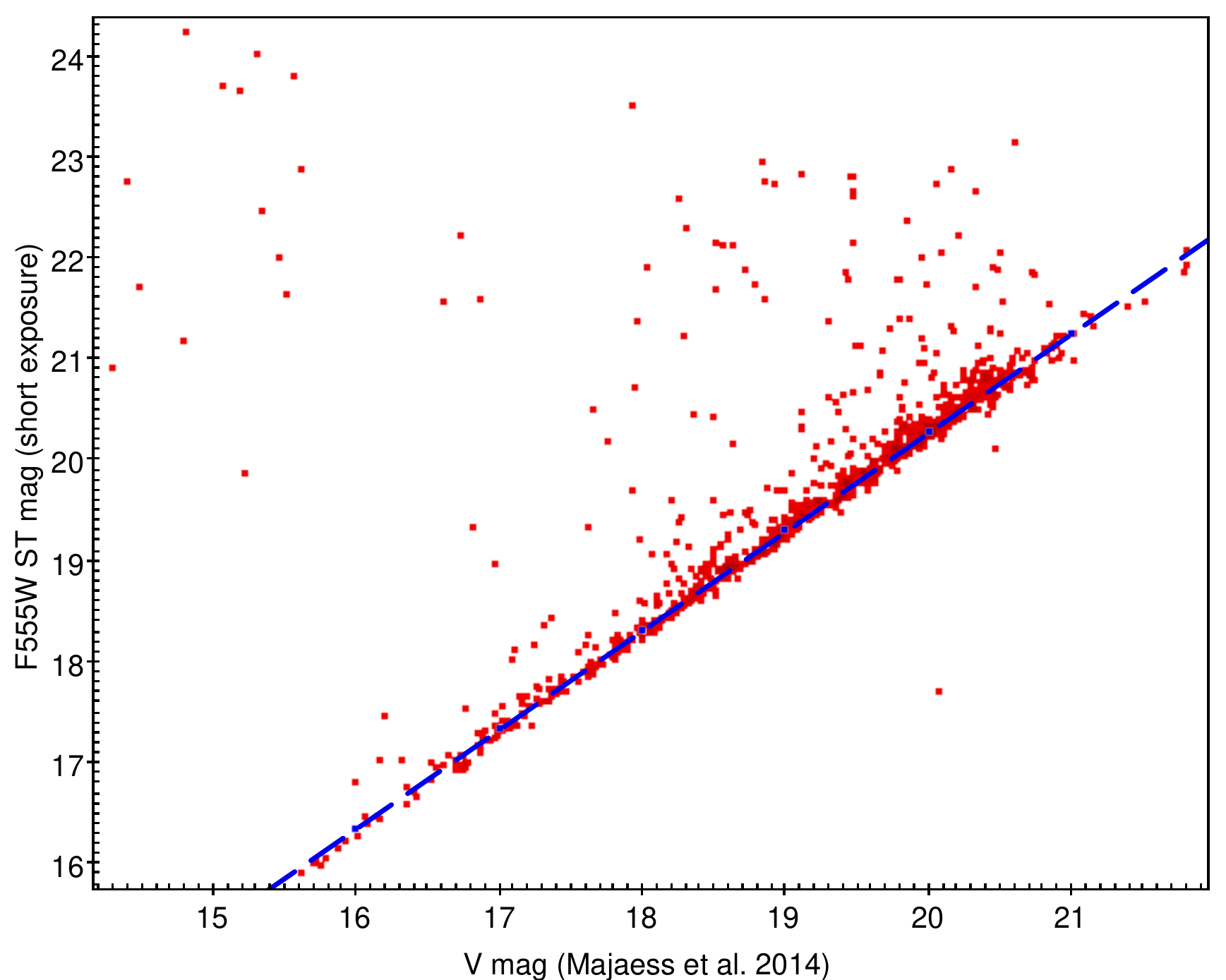}
\end{minipage}\qquad
\begin{minipage}[b]{.4\textwidth}
\includegraphics[width=\textwidth]{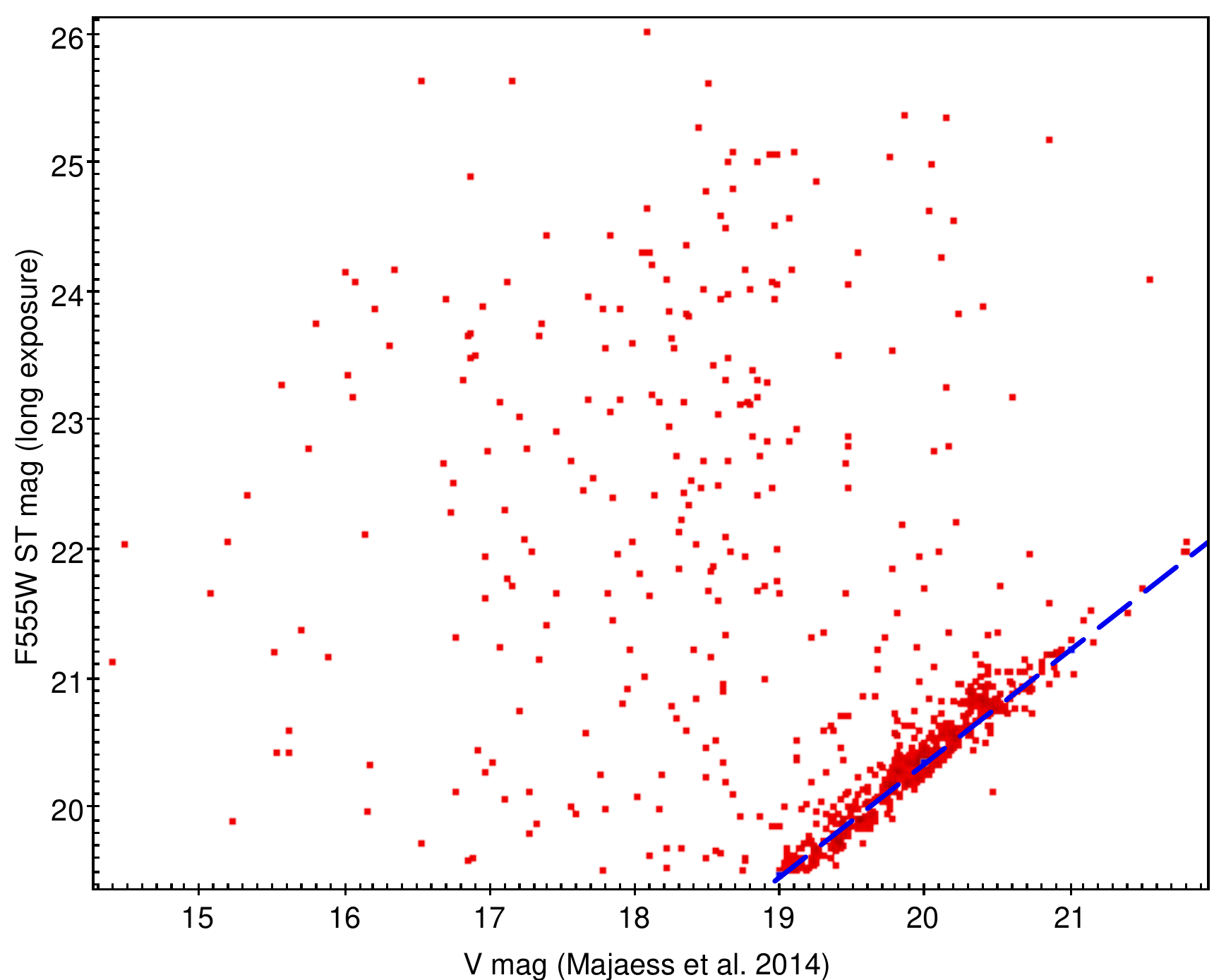}
\end{minipage}
\caption{Comparison of our measured F555W short exposure (left panel) and long exposure (right panel) magnitudes with the V magnitudes from Majaess et al. (2014). The apparent scatter is due to star lending issues caused by the relatively low resolution of ground based observations towards a crowded field. As the conversion of the HST F555W filter depends on the SED of each star a least square fitting routine was applied to both the short and long HST exposures. After removing  outliers the solutions are visualized with the blue tracks. Both fits are very similar with our short exposure data give a solution of F555W=0.98V+0.68, while the long exposure data give a solution of F555W=0.88V+2.75. As our short exposure data have a better overlap with the data of Majaess et al. (2014) we expect the solution from our short exposure data to better represent the relation between the different filters.}
\label{fig2}
\end{figure*}

\subsection{Constraining the physical parameters of AL~1}

A deep F555W-F814W versus F555W CMD was constructed for AL~1 using our derived F555W and F814W VEGA magnitudes for 5118 stars in our field of view (Fig. 3). The cluster lies at low Galactic latitude and as a result suffers from high interstellar extinction and strong contamination by field stars (see Carraro et al. 1995). This is evident from our CMD and makes difficult the recognition of the cluster's TO point.

\begin{figure*}
\centering
\includegraphics[scale=0.6]{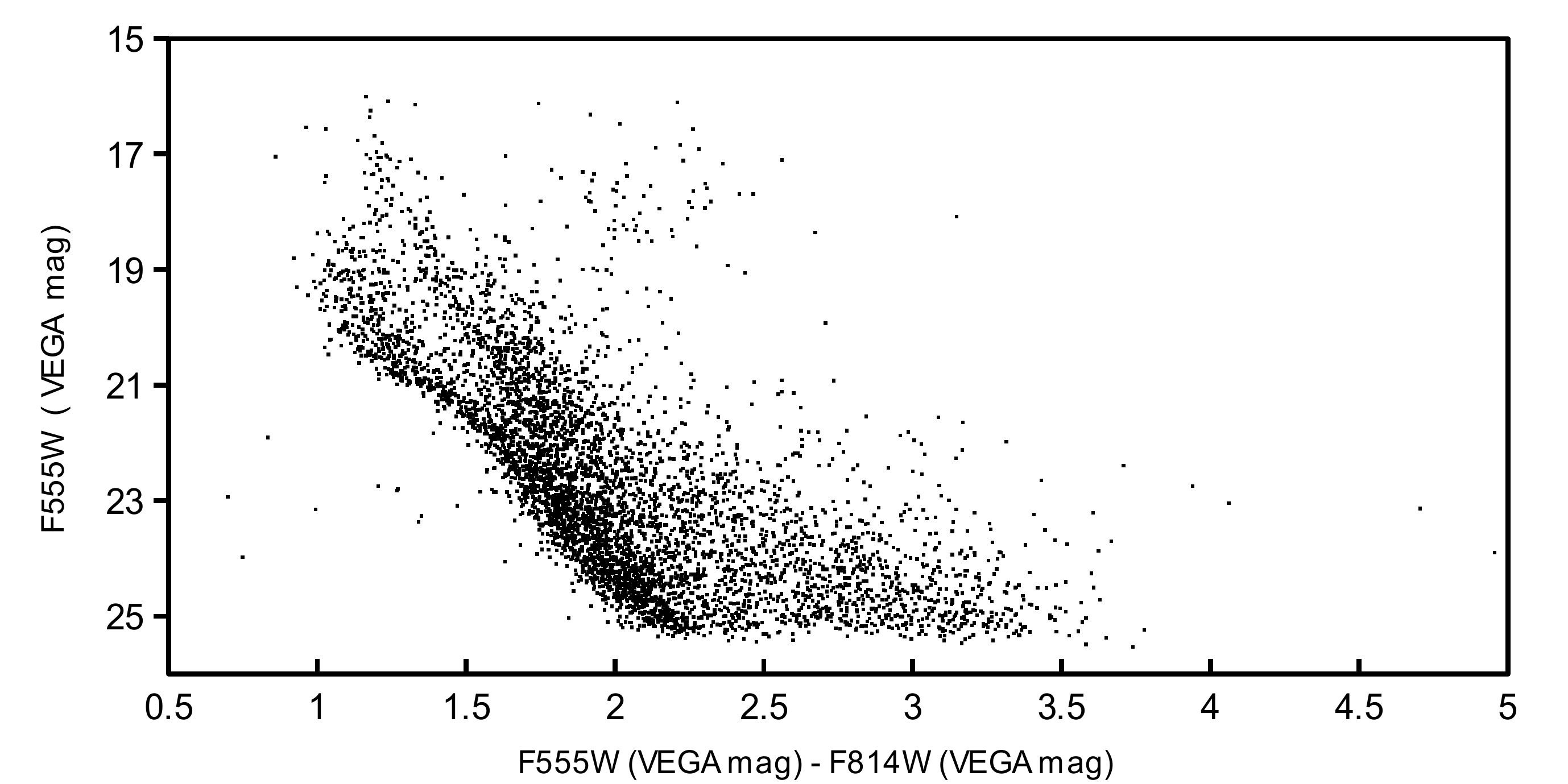}
\caption{The AL~1 cluster F555W-F814W versus F555W CMD from the measured VEGA magnitudes of all (non saturated) stars in our HST field of view. The main 
sequence of the cluster can been seen as a tight locus to the left, distinct from that of the field stars.}
\label{fig3}
\end{figure*}

For this reason, the Bayesian field star decontamination algorithm from the ASTECA code (Perren, Vasquez \& Piatti 2015) was used. After rejecting stars 
with magnitudes that have very large errors, it uses photometric data to determine the cluster's radial density profile and radius and assigns cluster membership 
probabilities (for a complete description of the code see Perren et al. 2015). The cluster's centre was located around RA= 13:15:16 and DEC=-65:55:16 and its 
radius estimated at 57~arcseconds. 
The code provided 1180 stars with membership probabilities larger than 50\%, which were consequently used for the construction of a 
CMD decontaminated from the majority of field stars (Fig. 4).

\begin{figure*}
\centering
\includegraphics[scale=0.5]{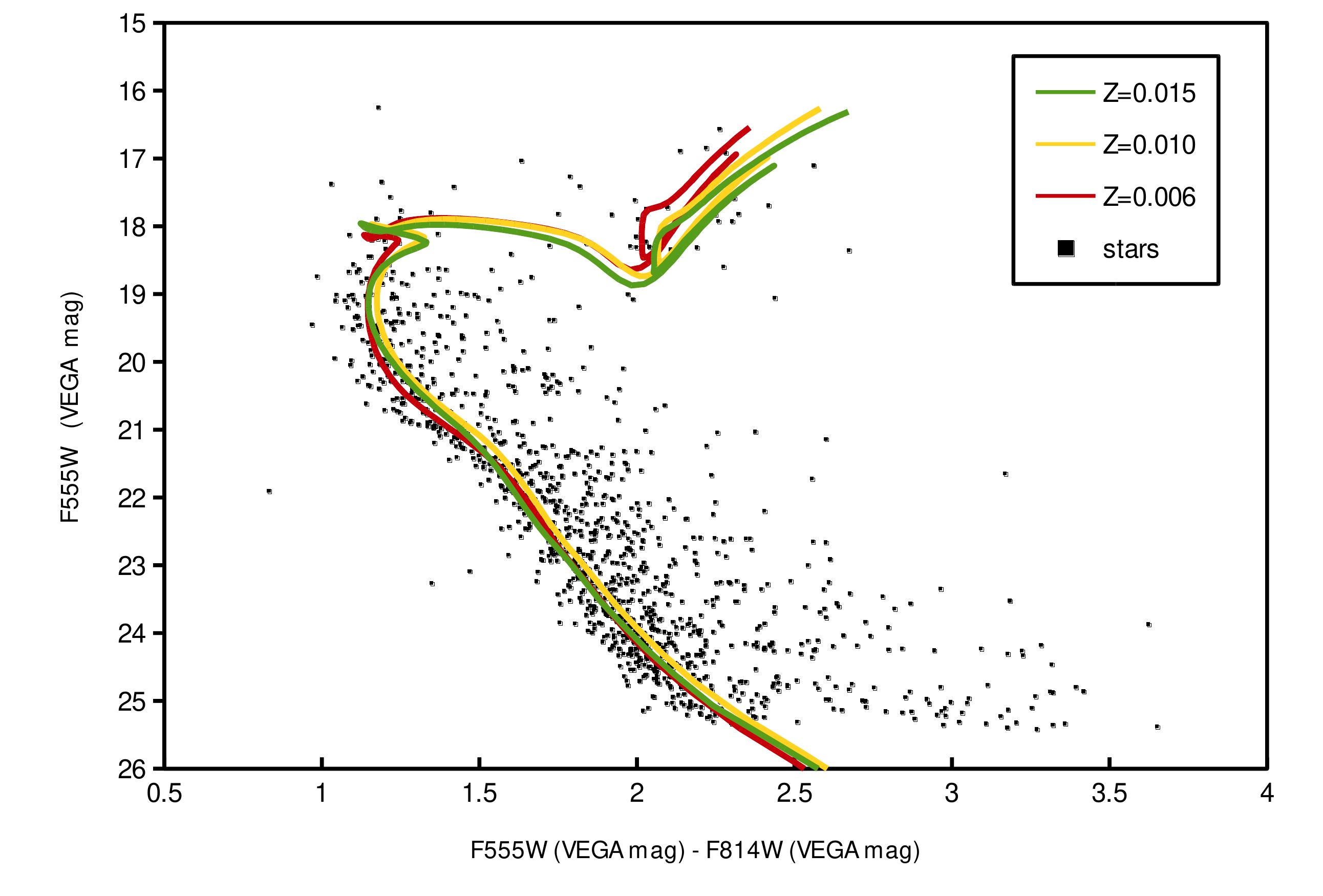}
\caption{Padova theoretical isochrones for different metallicities fitted to our  F555W-F814W versus F555W CMD after decontamination by field stars. The mean F555W and F555W-F814W errors of our stellar data are 0.08 and 0.09 respectively.  The parameters that produced each isochrone are for Z=0.015: age=0.66 Gyrs, E(B-V)=0.77, distance modulus=15.6, for Z=0.01: age=0.66 Gyrs, E(B-V)=0.82, distance modulus=15.4 and for Z=0.006: age=0.66 Gyrs, E(B-V)=0.83, distance modulus=15.4 .}
\label{fig4}
\end{figure*}

Our new improved cluster CMD's main features can now easily be seen.
The TO point is located around F555W= 18.6 VEGAMAG and F555W-
F814W= 1.12, while the main sequence ends around F555W= 18.1 VEGAMAG and F555W-F814W= 1.09. The red giant clump can be seen between F555W= 
17.6 -18.0 VEGAMAG. 

Although the ASTECA code can also be used for deriving a cluster's physical parameters, it is not reliable for relatively young clusters that suffer from strong field 
contamination (Perren et al. 2015) as is the case here. Since the brightest stars in our field are saturated, there is a lack of evolved stars in our 
data. As such we decided not to use the ASTECA code for this purpose.  Instead, as a starting point for the cluster's age derivation, the age index $\Delta$V was used as 
described by Carraro \& Chiosi (1994), which is defined as the difference between the V magnitude of the red clump and 0.25 magnitudes below the end of the 
MS. The age index $\Delta$V is then linked to the age ($\tau$) of a cluster as

\[
log(\tau_{Gyrs})=0.45 (\pm0.04) \times \Delta V + 8.58 (\pm0.23) \tag{1} \label{Eq. (1)}
\]

\medskip
\noindent
(Carraro \& Chiosi 1994, their equation 3). In our case and using the mean magnitude of the red clump (F555W=17.8), the age index is around 0.55, which translates in an age of 0.68 $_{-0.30}^{+0.54}$ Gyrs, which is in good 
agreement with previous estimates (see Table 1). The age spread is given from the terms in parentheses in Equation 1 as it is larger than the error induced by the red clump spread.

The reddening estimation was made by visual fitting of our CMD to theoretical Padova isochrones (Girardi et al. 2000; Bressan et al. 2012; Chen et al. 2015) for 
the ages close to that derived above and with different metallicities (see Fig 4). A best fit isochrone is found for a metallicity of Z=0.006, an age of 0.66 Gyrs and a 
colour excess of E(B-V)  $\approx$ 0.83 $\pm$ 0.05 ($A_{\rm V}$ = 2.57 $\pm$ 0.16, using the extinction laws by Cardelli, Clayton \& Mathis 1989 and $R_V$=3.1) 
that reproduces 
the main features of our data (see Fig. 5). From this reddening value and the location of the TO point we derive a distance modulus (m - M) = 15.4 $\pm$ 0.1, 
which translates into a cluster distance of 12 $\pm$ 0.5 kpc. Among the results from previous authors (see Table 1) this distance is only consistent with the distance 
values given from Carraro et al. (1995) and Carraro \& Munari (2004). The errors reflect the visual fitting of the theoretical isochrone (maximum error is the point 
where it is clear that the theoretical isochrone no longer fits our data). The constrained physical parameters for AL~1 imply a turn off mass for the cluster's stars of 
around 2.2$M_\odot$.

\begin{figure*}
\centering
\includegraphics[scale=0.6]{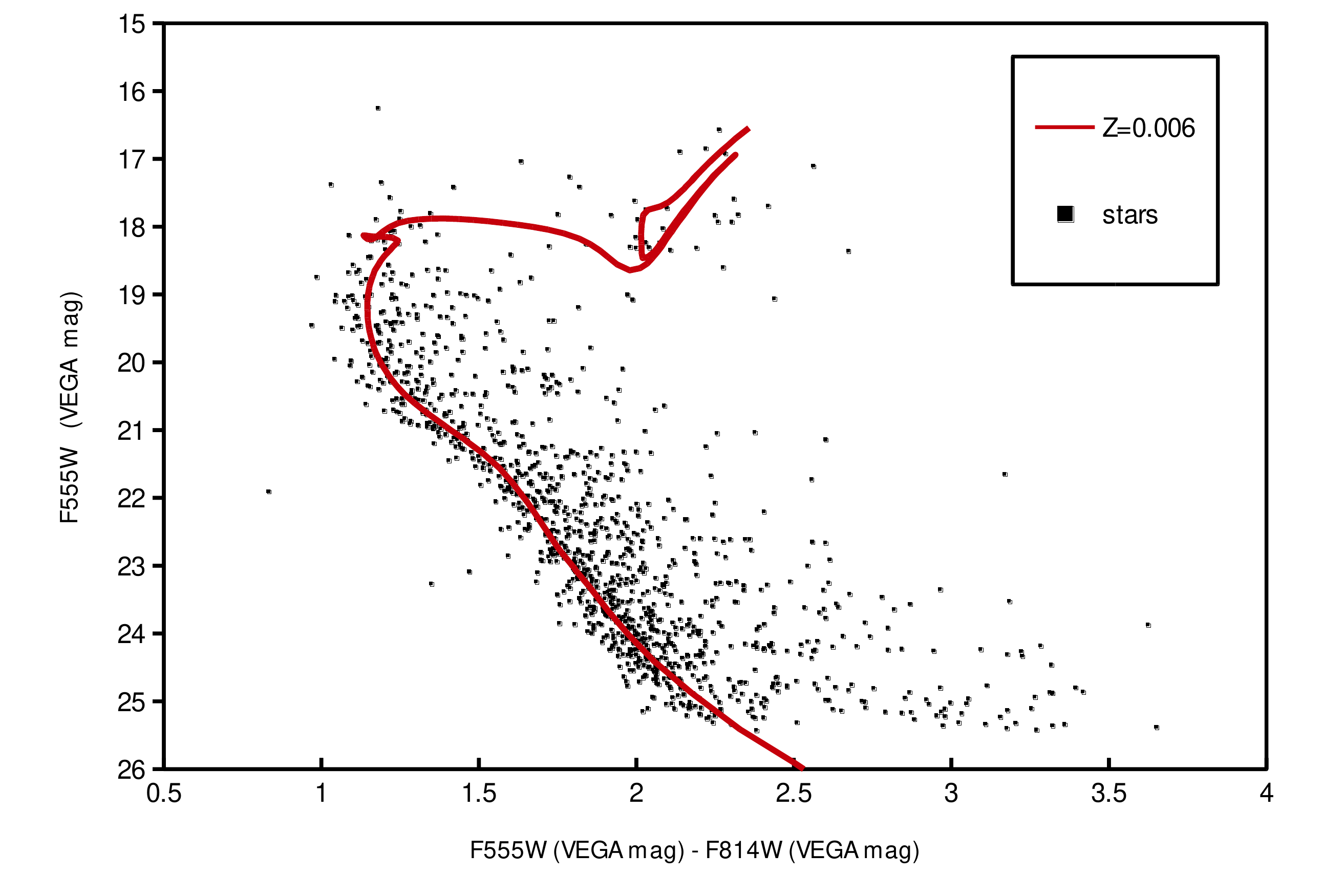}
\caption{The best fit Padova theoretical isochrone fitted to our F555W-F814W vs F555W CMD. Adopted parameters are as in Fig. 4 for Z=0.006.}
\label{fig5}
\end{figure*}

\section{The Bipolar Planetary Nebula PHR~1315-6555 and its CSPN}

One of the main motivations for our HST observations was the clear identification of the CSPN, which was not possible from our previous data. The sensitivity and 
spatial resolution of HST enables the detection and measurement of the faint CSPN of PHR~1315-6555 against its background nebula in the crowded field of 
AL~1 for the first time.

\subsection{Observations and analysis}

Our long F555W and F814W exposures allow the faint CSPN to be resolved and the determination of its V and I band continuum. The effect of binarity has been 
proposed as a possible explanation for the formation of bipolar PNe (De Marco 2009) as it is clear from the HST imagery that  PHR~1315-6555 is a bipolar PN. 
Our deep F814W band exposure was thus carefully examined to investigate the possibility of a cool companion. 

Additional HST exposures were obtained with the narrow-band [O~III] WFC3 F502N filter to show the PN in finer detail. 
These yielded a signal-to-noise around 10 per resolution 
element. The long-pass F200LP and F350LP filters give a signal-to-noise around 20 for a CSPN with the expected properties. The F200LP filter collects light of all 
wavelengths where the detectors are sensitive presenting a remarkable sensitivity at near-UltraViolet (near-UV) wavelengths (Dressel 2017), while the F350LP 
filter passes all visible light blocking the UV. As a consequence the difference of its calibrated flux from  those of F200LP gives the near-UV continuum (see 
Moreno-Ibanez et al. 2016). The F502N filter exposure reveals the nebular microstructure (Fig. 6) confirming its bipolar morphology, while the UV continuum 
ensures that the ionising CSPN will be correctly identified as it is expected to be the brightest in this bandpass. The images have been processed as described in 
section 2.1. 

\subsection{The Physical Parameters of PHR~1315-6555}

This unique PN, in terms of its confirmed location in a Galactic Open cluster with accurately know distance and progenitor mass, rests only 23 arcseconds from the cluster's centre. It was 
previously reported that it has an ionised mass of 0.5~$M_\odot$ (Parker et al. 2011), though in this work we found a smaller value (see later). Furthermore, its optical image indicates that it is evolved and probably optically thick (Parker et al. 2011). The presence of a strong HeII 4686 \AA\ emission line shows that it is a high excitation nebula. The crossover (Ambartsumyan) method (Kaler \& Jacoby 1989) predicts a CS apparent visual magnitude of 23.5~$\pm$~1 and effective temperature of 20.9 $\times$ $10^4$K (Parker et al. 2011). The calculated PN excitation class parameter Exp = 9.8 (Reid \& Parker 2010) predicts an even higher CS effective temperature, around 26.5 $\times10^4$ K, but such high CS temperatures are not expected for such evolved PNe (Parker et al. 2011 estimated a PN age around 11,000 yrs) and a temperature of 10 $\times$ $10^4$ - 14 $\times$ $10^4$ K and CS mass of around 0.6 - 0.65$M_\odot$ seems more reasonable (Parker et al. 2011).

Adjusting the contrast in the [O~III] narrowband image (see Fig. 6, top right), the full extent of the nebula is revealed allowing the measurement of its apparent 
diameter. The nebula has two main lobes with a NW-SE oriented waist, with some faint emission in its SE side. Drawing two axes, along the nebular waist (length 
4.48~arcseconds) and lobes, we estimate the nebular apparent diameter as the length of its major axis of 14.3~arcseconds. An adopted distance of 12 $\pm$ 0.5 kpc results in a PN physical diameter of 0.83 $\pm$ 0.04 pc.

\begin{figure*}
\centering
\includegraphics[scale=0.6]{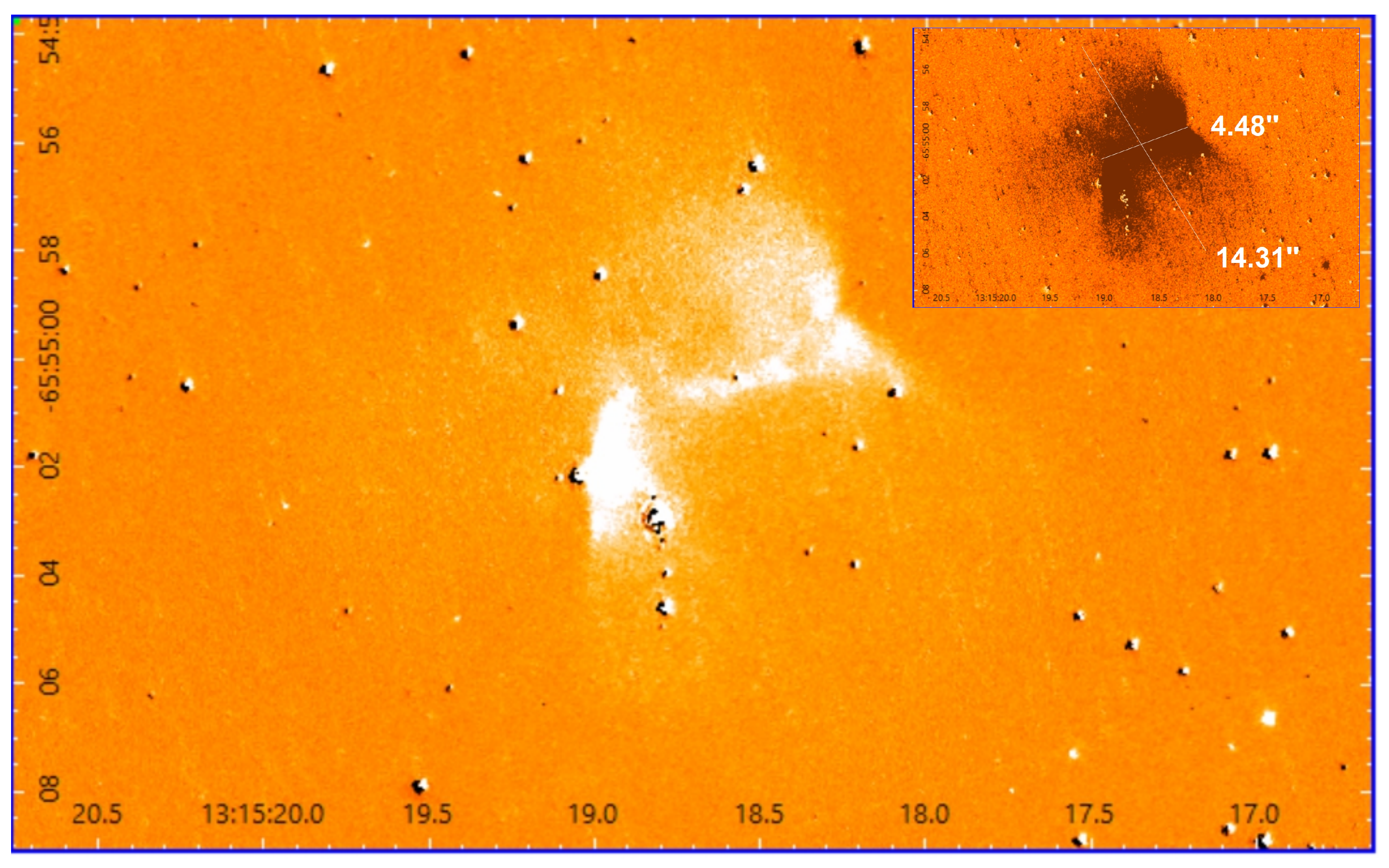}
\caption{The visual continuum (F555W) subtracted, flux calibrated [O~III] (F502N) nebular image. North is on the top, East on the left. The full extent of the clearly bipolar nebula can been seen on the top right, adjusting the contrast on the main image.}
\label{fig6}
\end{figure*}

\subsection{Identifying the CSPN of PHR~1315-6555}

For identifying our CSPN, the F200LP and F350LP filter exposures were used since their subtraction can reveal the bluest star in the nebular field. Using the 
IRAF/DAOPHOT package (Stetson 1987; Davis 1994) and the same procedures as before for the determination of the F200LP and F350LP VEGA magnitudes 
(see Section 3.1), a F200LP-F350LP versus F200LP CMD of all unsaturated stars in our field of view was constructed for locating the expected F200LP-F350LP 
colour of the relatively blue stars in our field. A Z=0.006 subsolar Padova theoretical isochrone for the cluster parameters derived in the previous section was fitted to our data 
(Fig. 7) to locate the main sequence. The small shift of our data to the red can be explained by their mean F200LP-F350LP error of 0.068. All blue stars in the field, 
including our CSPN, should be located on the left of the main sequence.

\begin{figure*}
\centering
\includegraphics[scale=0.65]{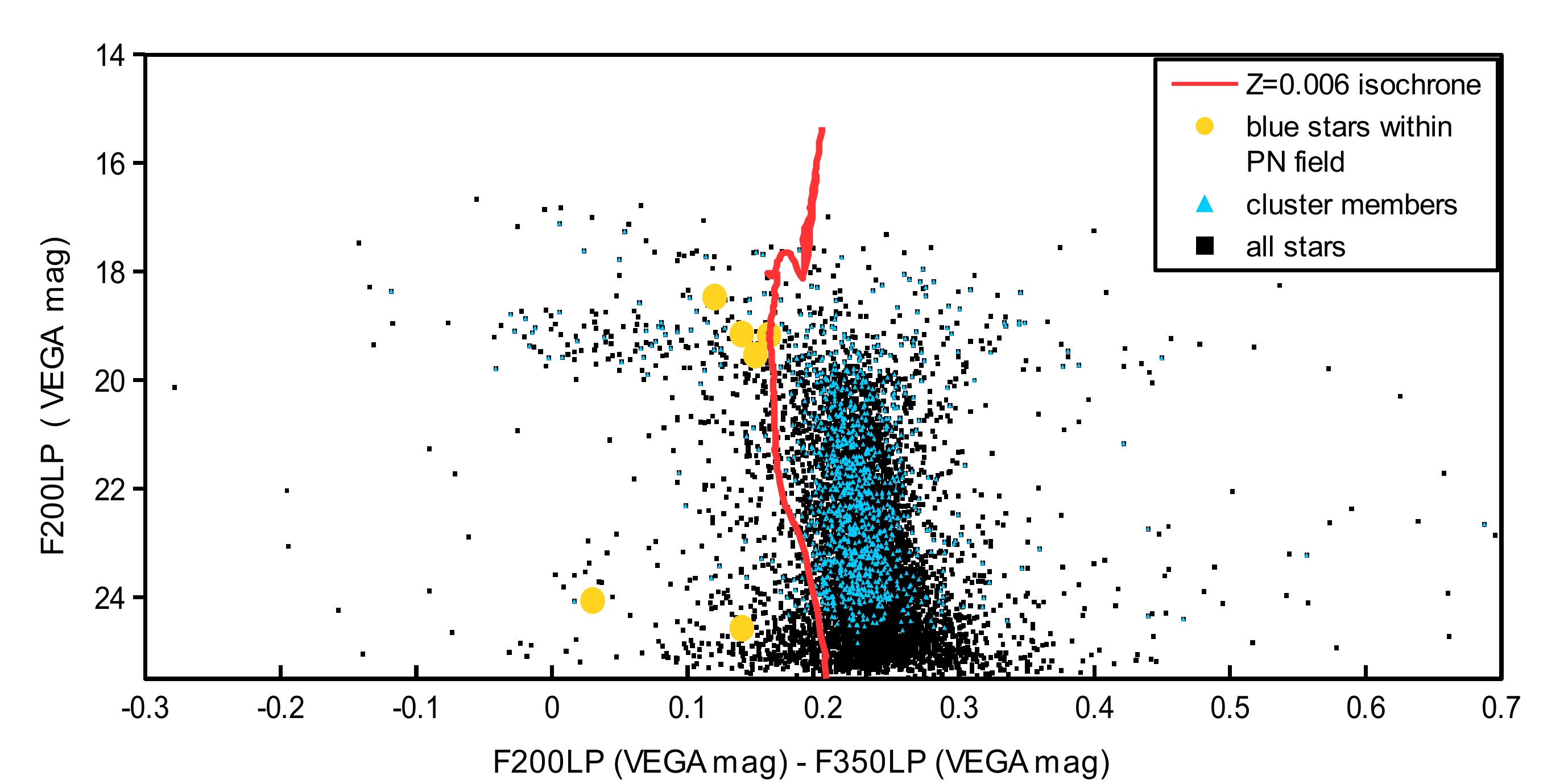}
\caption{A subsolar Padova theoretical isochrone for the cluster parameters derived before fitted to our F200LP-F350LP versus F200LP CMD from the measured VEGA magnitudes of all unsaturated stars in our field of view. The small shift of our data to the red can be explained by their mean F200LP-F350LP error of 0.068. The blue triangles indicate the cluster members as derived before by the decontamination algorithm from 
the ASTECA code (Perren, Vasquez \& Piatti 2015). The yellow circles indicate the six stars within the nebular field that lie blue-wards of the main sequence.}
\label{fig7}
\end{figure*}

Only 6 stars in the nebular field lie blue-wards of the main sequence (towards small F200LP-F350LP colours, see Fig. 7), which are indicated in circles on the F555W flux 
calibrated long exposure nebular image (Fig. 8). The star that is located closest (just South) of the centre of the nebular waist shows a reddened F200LP-F350LP= 
0.32 $\pm$ 0.08 indicating that is actually relatively red and thus, cannot be our CS. As the bluest star in the field with RA: 13:15:18.72 and DEC: -65:55:01.16 (circled in red in Fig. 8) lies only 1.46~arcseconds from the centre of the nebular waist (that has an apparent diameter of 4.48~arcseconds and is also indicated in Fig. 8), we are confident that this is the true CSPN.

\begin{figure*}
\centering
\includegraphics[scale=0.85]{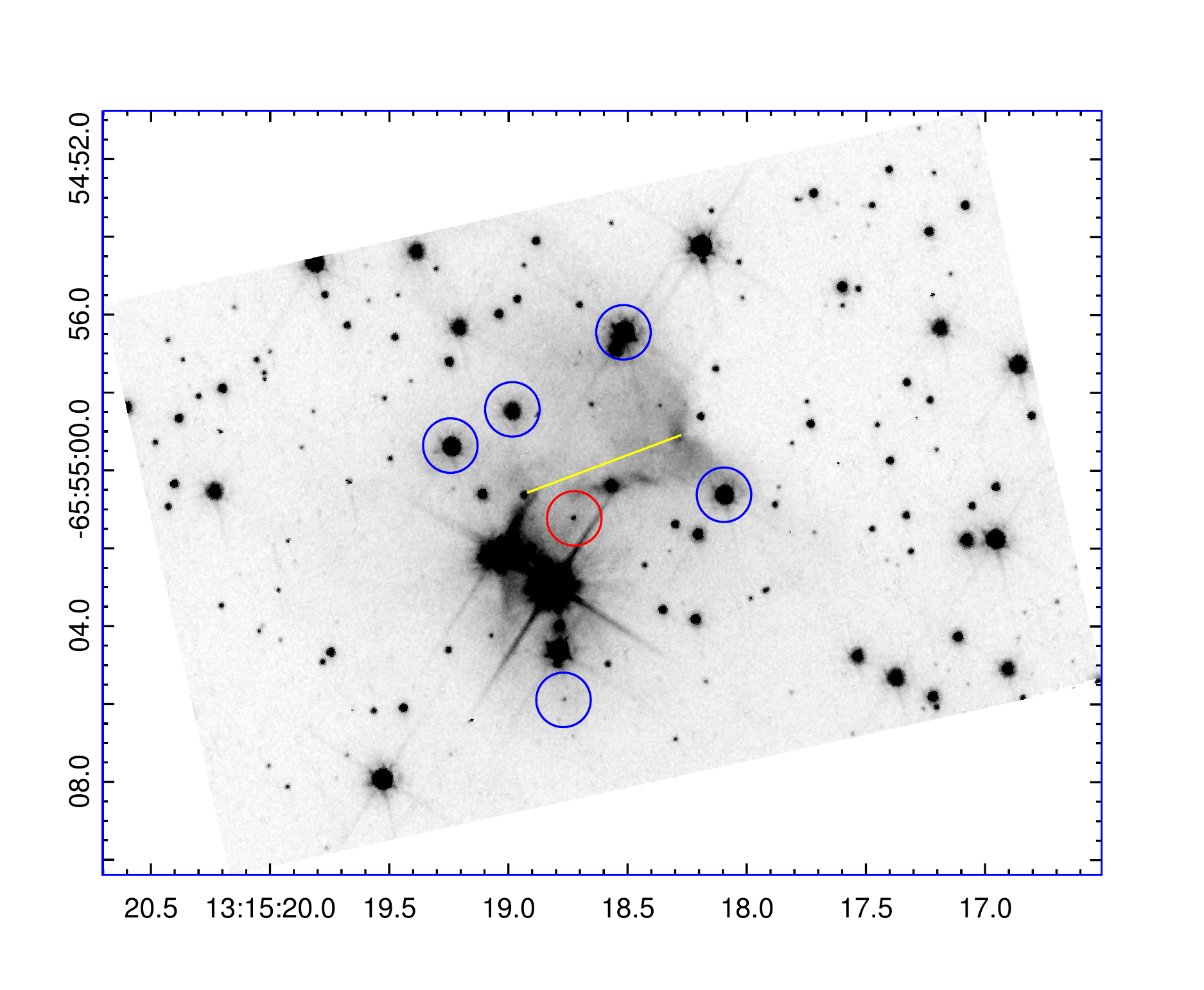}
\caption{The F555W flux calibrated long exposure nebular image. The blue stars are indicated by circles. The bluest star of all in the nebular field, lies only 
1.46~arcseconds from the centre of the nebular waist and is circled in red. The yellow line indicates the nebular waist that has an apparent diameter of 
4.48~arcseconds. North is on the top, 
East on the left. }
\label{fig8}
\end{figure*}

\subsubsection{Aperture photometry}

The newly identified CSPN F555W, F814E, F200LP and F350LP VEGA magnitudes were measured by aperture photometry using the IRAF/PHOT task (Davis 
1989). For the nebular subtraction we decided not to use our monochromatic [O~III] image as this approach induces large errors in the derived CSPN magnitudes. 
Instead, after determining the stellar radial profiles, we measured the flux inside a circular aperture around the stars' centre subtracting the background and 
nebular contribution by selecting an annulus close to the selected apertures enough to sample it. The selected width of sky annulus of 3 pixels (just outside our 4 
pixel aperture) ensures that the nebular continuum will be suitably subtracted. This approach may affect our photometric measurements as some small portion of 
the stellar PSF may fall outside our selected apertures. We consider the effect negligible, as the background will be still dominated by the nebular flux (see 
Moreno-Ibanez et al. 2016). 

The measured stellar fluxes are then transformed to ST and then VEGA magnitudes (see Section 2.1). Uncertainties reflect the systematic errors in the flux 
derivation, such as the deviation of the RMS of the background about its median value and the Poisson uncertainty in the flux measurements (see Moreno-Ibanez 
et al. 2016).

\subsubsection{Extinction Correction}

No internal nebular extinction could be measured from the 
spectral data of Parker et al. (2011) as is usually the case for such evolved and faint PNe. Interstellar extinction uncertainties are rarely a problem in optical 
wavelengths when the flux calibration is accurate as in their work. In the following we assume that there is no nebular internal extinction present. Our magnitudes were corrected for interstellar extinction following Kaler \& Lutz (1985) and using the E(B-V) = 0.83 $\pm$ 0.05 reddening value derived above 
(see Section 2.2), which agrees with the PN reddening that Parker et al. (2011) found from the Balmer decremant (E(B-V)=0.83 $\pm$ 0.08). The adopted reddening value affects the transformation of our measured VEGA magnitudes to the standard Johnson-Cousins System and may 
lead to a larger error (Moreno-Ibanez et al. 2016). The resulting dereddened F555W, F814W, F200LP and F350LP VEGAMAG are presented in Table 3.

\subsubsection{Transformation of Derived VEGA magnitudes to the Johnson-Cousins Magnitude System }

Our derived F555W VEGAMAG may not be adequate for calculating the effective temperature and luminosity of our CSPN and transformation to the standard 
Johnson-Cousins system is essential. The transformation depends on the Spectral Energy Distribution (SED) of our object. A black body spectrum can be 
regarded as a good approximation for representing the SEDs of CSPNe (Gabler, Kudritzki \& Mendez 1991).

Following the same steps as in Moreno-Ibanez et al. (2016), the required colour was estimated by synthetic photometry using the IRAF/STSDAS SYNPHOT 
package, which assumes a black body spectrum for representing an object's SED and calculates the difference in the magnitude between the two systems (for a 
complete guide of the SYNPHOT package see Laidler et al. 2005). For the calculation of our CSPN's Johnson V and I magnitudes we used our derived F555W and F814W VEGA magnitudes respectively as the passbands of these two HST filters are the most similar to those of the standard V and I magnitudes. Since we do not know the CS effective temperature in advance, we calculated 
their V-F555W difference taking the median of colours derived assuming black body spectra of effective temperatures between 30000 K and 300000 K in steps of 
5000~K adding their standard deviation quadratically to the errors of the derived Johnston magnitudes. The lower limit used for the effective temperature of a CSPN is the lower temperature required for the production of adequate ionizing photons to form a PN, while the upper limit is the higher CSPN temperature suggested by the Vassiliadis \& Wood (1994) evolutionary tracks.

After estimating the effective temperature of our CSPN from the HeII nebular emission line flux (see Section 3.3.4) and the derived V magnitude we repeated our 
transformation calculations using this fixed temperature, recalculated the V magnitude and iterated this process till reaching convergence (derived effective 
temperature variations less than 10 K and stable transformation colour). The final derived effective temperatures have also been used for the transformation of our 
measured F814W to Johnston I VEGAMAG. The calculated final standard V and I VEGAMAG along their errors are also presented in Table 3.

\begin{table} 
\center
\caption[]{The extinction corrected F555W, F814W, F200LP, F350LP and Johnson V and I VEGA magnitudes of our CSPN, as measured by aperture photometry, 
along with their errors.}  
\label{table3}

\begin{tabular}{cc}  
\noalign{\smallskip}  

\hline
\hline

Filter & VEGAMAG \\
\hline
F555W & 21.90 $\pm$ 0.60 \\
F814W &  22.65 $\pm$ 0.75 \\
F200LP &  20.73 $\pm$ 0.37 \\
F350LP & 21.46 $\pm$ 0.36 \\
V & 21.82 $\pm$ 0.60 \\
 I  & 22.65 $\pm$ 0.75 \\

\hline

\end{tabular}
\end{table}

\subsubsection{Zanstra Temperatures}

The CSPN effective temperature has been calculated in Pyneb (Luridiana, Morisset \& Shaw, 1995) via the well-known Zanstra method, described by Zanstra 
(1931) and developed by Harman \& Seaton (1966) and Kaler (1983). The Zanstra technique assumes that the nebula absorbs all photons above the Lyman limit 
of H ($\lambda$ < 921 \AA ) or $He^+$ ($\lambda$ > 228 \AA). Comparing the flux of the H~I or He~II nebular recombination line with that of the stellar visual 
continuum enables the star's total ionizing flux to be determined. Each recombination gives a Balmer series photon as it is usually the case at high optical depth. 
The error of the CSPN temperatures obtained via this method is usually less than 30\% (see Gleizes, Acker \& Stenholm 1989) and is affected by the 
uncertainties in the measured fluxes (see Tylenda et al. 1989).

Under the so called Zanstra discrepancy (see e.g. Gruenwald \& Viegas 2002), the temperatures derived with this method from the He~II line ($T_{He}$) are 
usually higher and more accurate than those derived from the H~I line ($T_{H}$). For optically thin PNe, ($T_{H}$) generally underestimates the true CSPN 
effective temperatures. For high temperature values, the nebula is optically thick to the H ionizing radiation and their ratio (TR) approaches unity (e.g. Kaler 1983; 
Kaler \& Jacoby 1989; Gruenwald \& Viegas 2000). As the CSPN temperature, at which the nebular transition from optically thick to thin occurs, is positively 
related to the progenitor's mass a high Zanstra discrepancy usually implies low-mass progenitors (Villaver et al. 2002). In our case, the presence of He~I in the PN 
spectrum indicates that the nebula is optically thick, while that of He~II implies a high CSPN temperature (see Moreno-Ibanez 2016) so that a derived ($T_{H}$)  
should be considered a fairly good approximation to the true temperature value. However, the Zanstra temperature derived from the He~II line is preferred for the 
derivation of the candidate CSPN luminosities and masses as an even better approximation to their actual temperatures (see e.g. Villaver et al. 2003,2004; 
Moreno-Ibanez et al. 2016). We need to note the possibility of a nebula becoming optically thin to He~II at particularly high stellar temperatures (Shaw \& Kaler 
1989), in which case the Zanstra temperatures derived in this way should be considered as lower limits (see e.g. Villaver et al. 2003).

Using the H~I and He~II nebular spectral flux measurements by Parker et al. (2011), corrected for interstellar extinction (see Section 3.3.2.) using the reddening 
value derived above and our measured extinction corrected stellar visual magnitude we calculated both the H~I and He~II Zanstra temperatures, which are 
presented along their ratio (TR) and errors in Table 4. The temperature errors reflect uncertainties of 15\% (see Preite-Martinez \& Pottasch 1983).

\subsubsection{Luminosities}

With the assumption that the Zanstra temperatures are good approximations for the CSPN effective temperature, we now calculate the Bolometric Correction 
factor (BC) using

\[
BC= 27.66-6.84 \times logT_{eff} \tag{2} \label{Eq. (2)}
\]
\smallskip

\noindent
(Vacca, Garmany \& Shull 1996), where for $T_{eff}$ we used our derived $T_{H}$ and $T_{He}$. This relation only slightly depends on a star's surface gravity 
though assumes a $T_{eff}$ of not more than 50,000 K (Vacca et al. 1996; Flower 1996), but considering that there are no other calibrations in the literature 
suitable for hot stars we assume the derived values to be valid for our purposes (see Moreno-Ibanez et al. 2016). 

The known PN distance allows the calculation of the absolute luminosity of our CSPN. Adopting a value of 15.4 for the distance modulus (see Section 2.2), we 
can estimate the CSPN absolute visual magnitude $M_v$ (=6.42) and adding the BC found previously we derive its luminosity `L' as

\[
L=-\frac{M_{V(bol)}-M_{\odot(bol)}}{2.5}log(L/L_\odot) \tag{3} \label{Eq. (3)}
\]
\smallskip

\noindent
where $M_{\odot(bol)}$= 4.75 mag (Allen 1976). The derived BC and luminosity are shown in Table 4, while their errors have been computed through standard error propagation.

\begin{table}
\center
\caption[]{The derived physical properties of our CSPN.}  
\label{table4}

\begin{tabular}{lcc}  
\noalign{\smallskip}  

\hline
\hline

$T_{H}$  Zanstra estimate & ($10^3$ K) & 69.03 $\pm$ 10.35 \\
$T_{He}$ Zanstra estimate &  ($10^3$ K) & 112.68 $\pm$ 16.90 \\
TR & &1.63 \\
\multirow{2}{*}{Bolometric Correction (BC)} &  (from  $T_{H}$) &  -5.44 $\pm$ 0.45 \\
  & (from  $T_{He}$) & -6.90 $\pm$ 0.45 \\
\multirow{2}{*}{log(L/$L_\odot$)} & (from  $T_{H}$) & 1.51 $\pm$ 0.30  \\
 & (from  $T_{He}$) & 2.09 $\pm$ 0.30  \\
\multirow{2}{*}{M/$M_\odot$ (VW tracks)} & (from  $T_{H}$) & < 0.50  \\
 & (from  $T_{He}$) & 0.62$_{-0.09}^{+0.12}$  \\
\multirow{2}{*}{M/$M_\odot$ (MBM tracks)} & (from  $T_{H}$) &  < 0.50  \\
 & (from  $T_{He}$) & 0.58$_{-0.08}^{+0.14}$  \\

\hline

\end{tabular}
\end{table}

\subsubsection{CSPN Masses}

The presence of the He~II line in a PN's emission spectrum implies effective temperatures larger than 50 $\times$ $10^3$ K (Osterbrock \& Ferland 2006). The 
effective temperature derived here from the H~I line is lower than predicted from the nebular spectral measurements (Parker et al. 2011), indicating that the nebula 
is optically thin relative to H~I. The temperature derived from the HeII line is still lower than what the PN's spectral data suggest (high He~II/H$\beta$ line ratio) but 
still consistent with what expected for an optically thick PN so better represents our data.

The CSPN H~I and He~II masses have been determined by plotting its derived luminosities and temperatures in the logT-logL plane along with the Vassiliadis \& 
Wood (1994, from now on VW tracks) and latest Miller Bertolami (2016, from now on MBM tracks) post-AGB evolutionary tracks for single stars and assume 
subsolar metallicities close to the cluster metallicity derived before (Z=0.008 for VW tracks and Z=0.01 for MBM tracks) shown in Fig. 9. VW evolutionary tracks are plotted for both H- burning and He- 
burning post-AGB stars, whose nature depends on the dominant burning shell at the time they leave the AGB phase. The He- burning tracks are more accurate 
for low-mass progenitors (Vassiliadis \& Wood 1994) but since He- burning CSPNe consist of only $\sim$25\% of the total CSPN population (i.e. Iben 1984; Wood 
\& Fulken 1986; Schoenberner 1986; Renzini 1989) and we account for an intermediate mass progenitor of around 2 $M_\odot$ (see Section 2.2) we derive the 
stellar masses (see Table 4) using only their H-burning tracks when possible (for the adopted metallicity VW H- burning trucks are not available for relatively low mass stars).

\begin{figure*}
\centering

\subfloat[]{%
  \includegraphics[width=\textwidth]{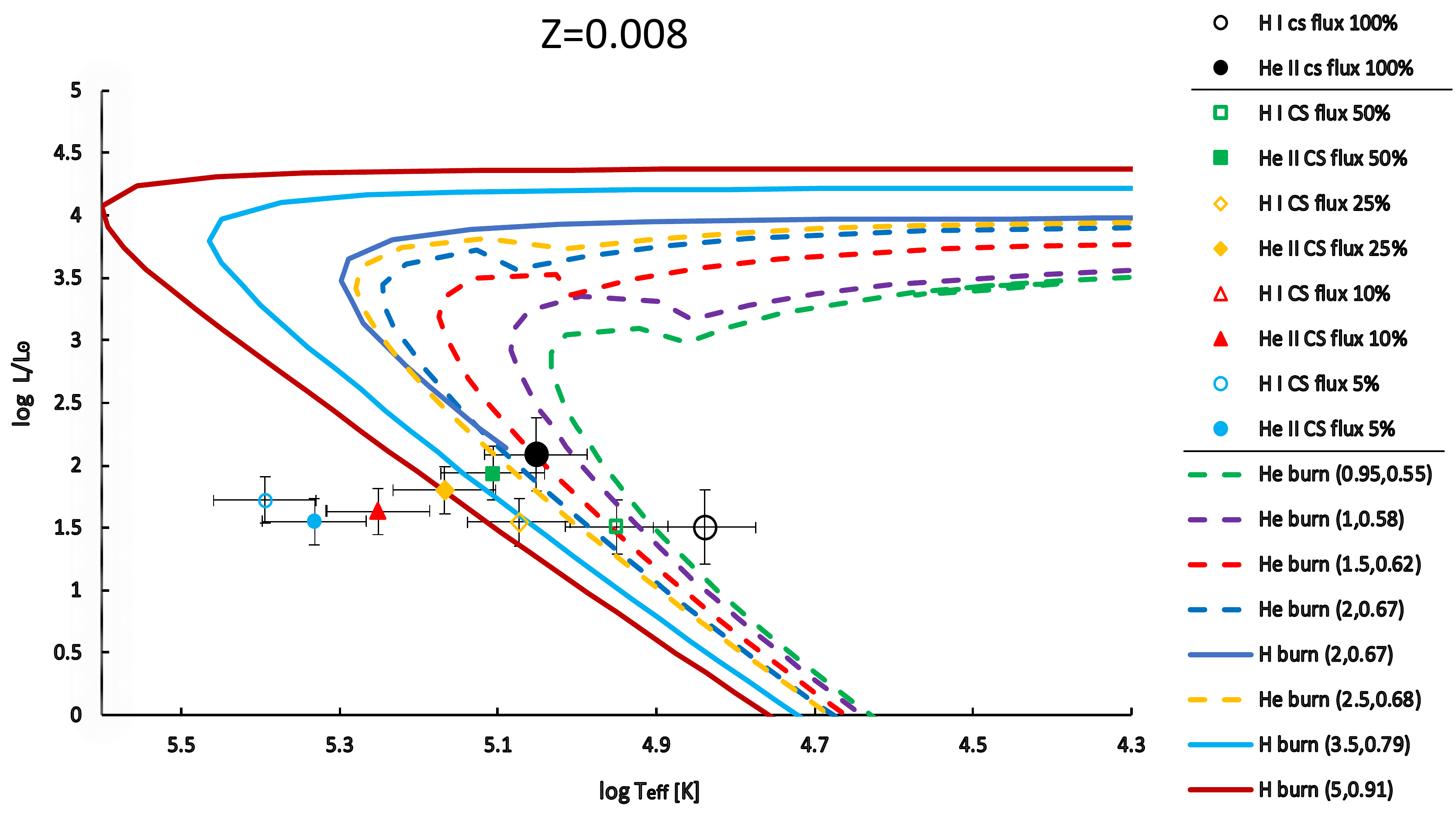}%
}

\subfloat[]{%
  \includegraphics[width=\textwidth]{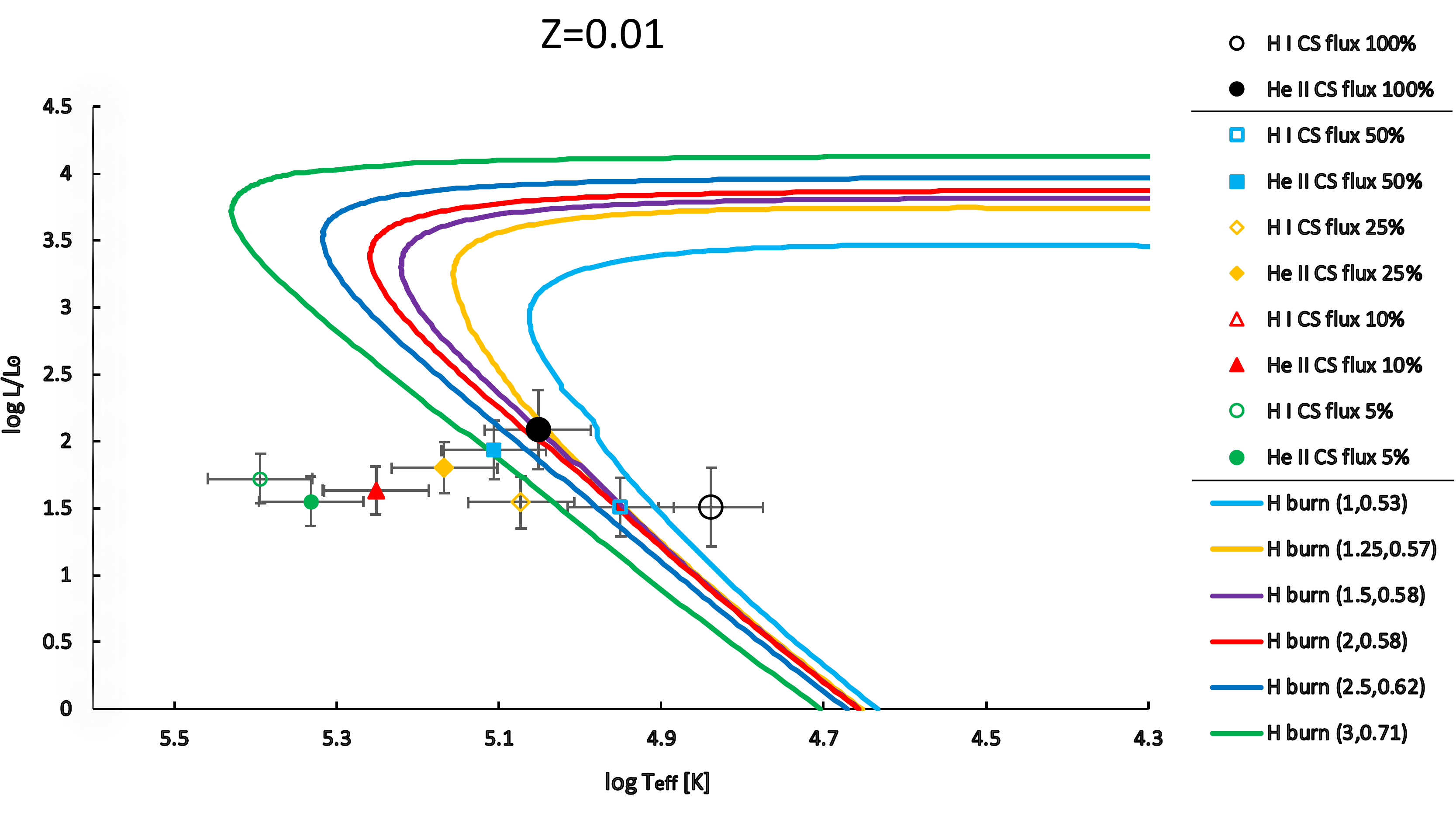}%
}

\caption{The computed physical parameters of our CSPN for different companion flux contributions plotted along with the VW Z=0.008 (upper panel) and MBM Z=0.01 (lower 
panel) evolutionary tracks. Open and full points represent the CS luminosities and temperatures calculated from Zanstra temperatures that were derived from the HI and He II PN emission lines respectively. The evolutionary tracks legend indicates the initial and final masses that produced them.}
\label{fig9}
\end{figure*}

The CSPN masses from both VW and MBM tracks, have been calculated by interpolating the values obtained from the closest tracks plotted reflecting in their 
errors the uncertainties in the luminosities and temperatures and are presented in Table 4. Both set of tracks well represent our data within the errors.

\subsection{Investigating the possibility of a stellar companion}

CSPN luminosities, effective temperatures and masses have been calculated assuming the lack of a binary companion. However, this is rather precarious 
particularly for a bipolar PN as complex PN shapes are believed to be the result of binary systems (De Marco 2009; De Marco et al. 2013; Garcia-Segura et al. 
2014). Its presence would imply that the derived effective temperatures and masses are only lower limits (Kaler 1983). 

For exploring the possibility of an unresolved companion to contribute to our measured visual flux and following the same procedure as in Villaver et al. (2004), 
we assumed that different portions of it are coming from an hypothetical stellar companion. We considered companion contributions to the total measured visual 
fluxes between 50\% and 95\%, as a smaller fraction is not expected to significantly change the CSPN mass estimates (Villaver et al. 2004). The cluster's known 
distance and reddening allows the determination of the mass and spectral type of the hypothetical companion (assuming that it is on the MS), from the cluster's 
theoretical subsolar isochrone adopted in Section 2.2. and its estimated visual magnitude (see De Marco et al. 2013, their table C1). Consequently, following the 
same steps as in Section 3.1, we calculated the new candidate CSPN visual magnitudes, effective temperatures, luminosities and masses, which are presented 
along with the corresponding derived companion masses and spectral types in Table 5. The quoted CSPN visual magnitudes, luminosities and masses were 
computed from the $T_{He}$ (c.f. Section 3.3.6) and either the H- or the He- burning evolutionary tracks (depending on their proximity to the CS points) and no masses were estimated for CSPN parameters that fall far from the 
evolutionary tracks as they cannot be reliable. The new CSPN physical parameters plotted on the logT-logL plane along with their initial estimates (assuming that 
100\% of the flux comes from the CSPN) and the subsolar metallicity evolutionary tracks from Vassiliadis \& Wood (1994) and Miller Bertolami (2016) are shown in 
Fig. 9.

\begin{table*}  
\caption[]{Derived parameters assuming different contributions from a binary companion.}  
\label{table5}
\centering
\begin{minipage}{20cm}
\begin{tabular}{l | cccc | cccccc}  
\noalign{\smallskip}  

\hline
\hline
\multicolumn{3}{c}{Companion Star} & &  \multicolumn{6}{c}{CSPN}\\  
\cline{1-3}
\cline{5-11}
& & & & & & & & & & \\
Contribution & M  &  Spectr. Type & & V\footnote{Corrected for interstellar extinction} & $T_H$ & $T_{He}$ &  log(L/$L_\odot$)\footnote{Derived from $T_{He}$} & VW M\footnote{Derived from $T_{He}$ and VW tracks} & MBM M\footnote{Derived from $T_{He}$ and MBM tracks} \\
(\%) & ($M_\odot$) & &  & (VEGAMAG) & ($10^3$) & ($10^3$) & & ($M_\odot$) & ($M_\odot$) \\
\hline
 50 & 0.66 & K7V & & 22.65 $\pm 0.33 $ & 89.15 $\pm 13.37$ & 127.77 $\pm 19.17 $ & 1.94 $\pm 0.22$ & 0.73$_{-0.13}^{+0.17}$ & 0.71 $\pm 0.14$  \\ 
 75 & 0.71 & K5V & & 22.21 $\pm 0.46 $ & 118.38 $\pm 17.76 $ & 147.06 $\pm 22.06 $ & 1.80 $\pm 0.19$ & 0.91$_{-0.19}^{+0.17}$ & $-$  \\ 
 90 & 0.74 & K5V & & 22.01 $\pm 0.54 $ & 178.39 $\pm 26.76$ & 178.47 $\pm 26.77$ & 1.63 $\pm 0.18$ & $-$ & $-$  \\ 
95 & 0.74 & K5V & & 21.95 $\pm 0.57$ & 247.91 $\pm 37.19 $ & 214.61 $\pm 32.19$ & 1.55 $\pm 0.18$ & $-$ & $-$ \\ 
\hline

\end{tabular}
\end{minipage}

\end{table*}

Assuming that no significant mass transfer occurs between two components of a binary system their mass difference should be large enough for allowing the 
more massive to evolve as a PN first (see Villaver et al. 2004), a condition that is satisfied in all cases explored here, where the possible companion is a K 
Dwarf. The PN distance implies that a Supergiant companion cannot be justified from the measured visual fluxes and Giant and Subdwarf O companions are 
unlikely due to their short lifetimes (Renzini \& Buzzoni 1986; Yungelson, Tutukov \& Livio 1993) and the short timescales during which they produce significant 
visual light (see Villaver et al. 2004) respectively. However, a MS, Subdwarf or Red Dwarf companion is still a possibility. From Fig. 9 we can see that a more than 
50\% flux contribution from a companion would result in a progenitor mass much larger than the cluster's turn off mass of $\sim$ 2.2$M_\odot$ so we can safely 
rule out this possibility. However, a companion contribution of $\leq$ 50\% to the total observed visual flux results in progenitor masses close to 2.2$M_\odot$
among the errors, and thus, such a case has to be further explored.

Our precisely measured F814W magnitude provides us hints regarding the possibility of the presence of an unresolved cool, low mass companion as a CSPN is 
expected to radiate primarily in UV and blue wavelengths and the companion will yield an infrared (IR) excess (see Moreno-Ibanez et al. 2016; Barker et al. 
2017). The observed IR excess is though not a definitive indication of a cooler companion and could instead imply an unusual stellar atmosphere or inaccurate 
reddening estimates (see Barker et al. 2017). Using the IRAF/STSDAS SYNPHOT package (see Section 3.1.3.) we used blackbody models and found the 
expected F555W-F814W colour for a CSPN temperature equal to that derived before assuming zero 
contribution from a companion. The adopted temperature value of 112,680 K reflects our yields of $T_{He}$. Any (reasonable for a CSPN) value would not 
significantly change our results. The modelled colours have been compared to our measured reddened F555W-F814W colours and are presented along with their 
differences in Table 6. As can be seen our CSPN present no IR excess, though the magnitude errors are quite large. This suggests that the presence of a cool companion is unlikely but cannot be excluded. Similarly, we calculated the expected F200LP-F350LP colour (presented in Table 6, along with its 
difference from our measured one) showing that the measured near-UV colour agrees, within the errors, with that expected for a CSPNe of the derived 
temperature.

\begin{table}  
\center
\caption[] {Derived data on any IR excess of our CSPN. Quoted magnitudes are in the VEGA magnitude system VEGAMAG and uncorrected for interstellar extinction.}  
\label{table6}

\begin{tabular}{lc}  
\noalign{\smallskip}  

\hline
\hline

 $T_{He}$ ($10^3$ K) & 112.68  $\pm$ 16.90\\
F555W& 24.47 $\pm$ 0.58 \\
F814W& 23.89 $\pm$ 0.75\\
F250LP & 24.28 $\pm$ 0.33 \\
F350LP & 24.19 $\pm$ 0.33 \\
\hline
 $(F555W-F814W)_{mod}$ & 0.749 \\
\hline
 $\Delta_{obs-mod}$(F55W-F814W) & -0.174  $\pm 0.665$ \\
\hline
$(F200LP-F350LP)_{mod}$ &   0.016 \\
\hline
$\Delta_{obs-mod}$(F200LP-F350LP)&  0.067 $\pm 0.463$ \\
\hline

\end{tabular}
\end{table}

\section{Discussion}

Our new HST provided CMD is a significant improvement on what was previously available for this cluster. It confirms previous studies that show that AL~1 is an 
intermediate-age, distant, highly reddened OC. A solar metallicity would not explain the high N/O abundance ratio of 0.87 present in the PN 
(see Parker et al. 2011) as the latest AGB models (Karakas 2010; Karakas \& Lugaro 2016; Ventura et al. 2018) predict that, in solar metallicity environments, such high N 
abundances cannot result from such low-mass stars. The adopted cluster metallicity of Z=0.006 marginally agrees with the latest predictions of AGB yields (Karakas \& Lugaro 2016) for the derived turn-off mass of $\sim$ 2.2$M_\odot$, though precise abundance studies will be required to clarify this issue, which are beyond the scope of this paper.

Our deep HST F200LP and F350LP filter images reveal that the nebular core is a faint blue star close to the centre of the bipolar waist (see Figure 8). It is unlikely 
that any other star could be the central star of the detected PN as it would either be unusually red or fall too far from the well resolved apparent nebular centre. It is also unlikely for an even 
fainter CSPN not to be visible at the defined distance with the depth achieved here, of V= 26 mag, unless it was a member of an unresolved binary system with a 
much brighter companion or hidden by a foreground star. This can be excluded based on our near-UV data, at least at the proximity of the nebular centre. An 
object fainter than V= 26 would be either much further than the cluster's adopted distance or unreasonably hot for a CSPN.

\begin{figure*}
\centering
\includegraphics[scale=0.6]{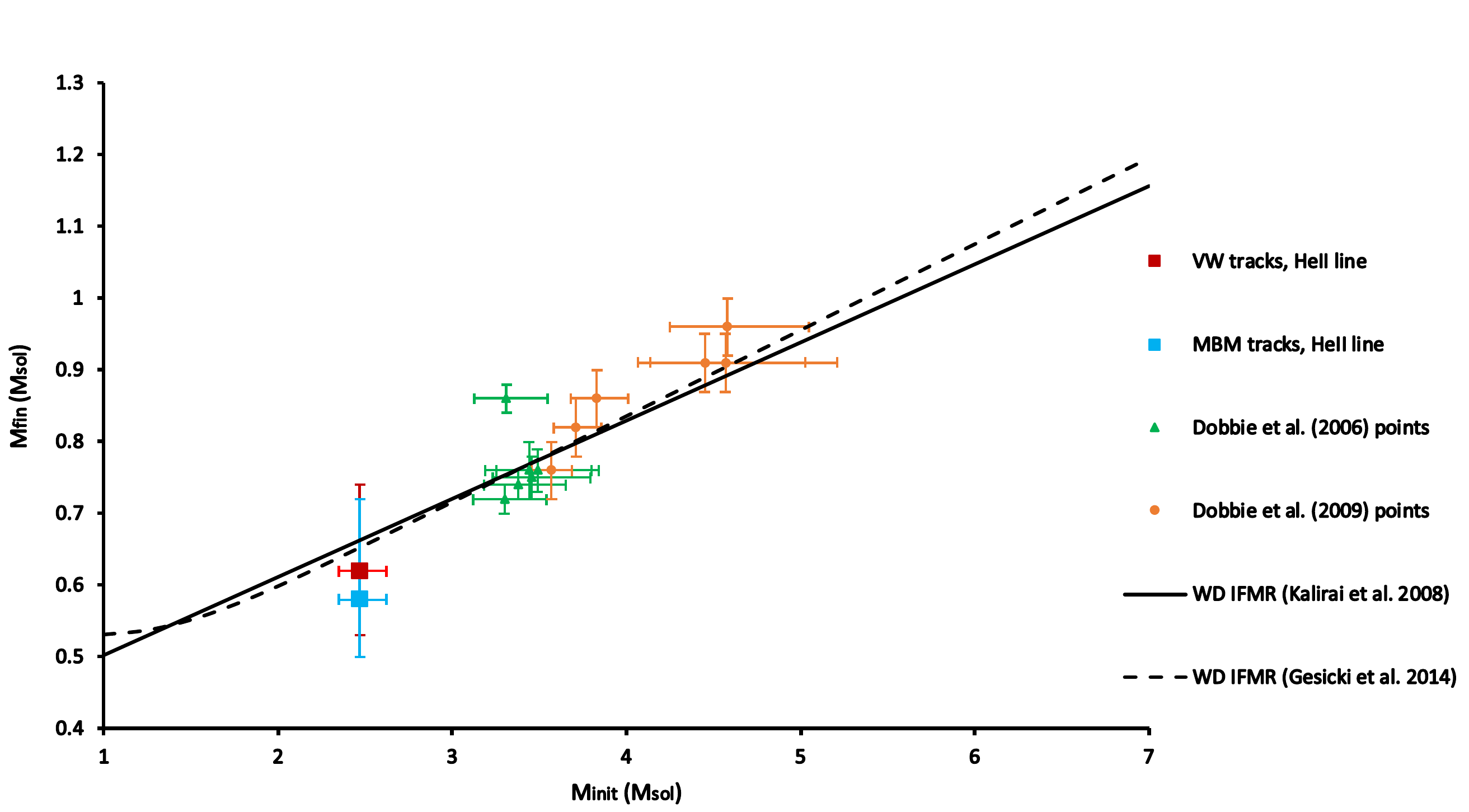}
\caption{ The computed CS initial (from the cluster's CMD) and final mass (from both the VW and MBM tracks) plotted along the initial and final mass points from Dobbie et al. (2006; 2009) and the Kalirai et al. (2008) and Gesicki et al. (2014) IFMR. The initial mass errorbars reflect the errors in the derived cluster parameters.}
\label{fig10}
\end{figure*}

Our results indicate that the true core of PHR~1315- 6555 is a faint CSPN, whose estimated parameters agree, within the errors, with the cluster turn-off mass and 
result in a PN core mass of around 0.62 $M_\odot$ from the VW tracks and around 0.58 $M_\odot$ from the new, faster, MBM tracks. This is about 
average for a Galactic CSPN (see Moreno-Ibanez et al. 2016) and consistent with what expected for a Type~I PN originating from a higher mass progenitor of $\sim$ 2.2$M_\odot$. 
The estimated effective temperature is still much lower than predicted from the crossover method (209 $\times$ $10^3$ K) and the excitation class parameter 
(265 $\times$ $10^3$ K, see Parker et al. 2011) but considering that both of these methods give just an approximation of the true effective temperatures our 
direct measurements supersedes these estimations. 

Theroretical cluster isochrones (Girardi et al. 2000; Bressan et al. 2012; Marigo et al. 2017) for the derived cluster parameters predict that the progenitor mass of post-AGB stars is around 2.47 $M_\odot$. The initial and remnant mass of our CSPN provide a unique additional point for the IFMR from WD studies. Figure 10 shows our results from both the VW and 
MBM tracks plotted along the initial and final mass points from Dobbie et al. (2006; 2009) and the WD IFMR from Kalirai et al. (2008) and Gesicki et al. (2014). As can been seen the final mass calculated from the VW tracks better fits their data.

The calculated PN intrinsic radius of $\sim$ 0.42 pc confirms the nebular evolved nature (Frew \& Parker 2010; Parker et al. 2011) and for a typical mean 
expansion velocity of 24 km/s (Frew 2008) its age is estimated around 17,000 yrs. The estimated nebular age also agrees within the errors with that predicted from the MBM 
tracks for a CS of the derived mass. The newly derived parameters for the PN distance and angular radius, and assuming a canonical spherical shape, 
a filling factor $\epsilon$= 1 and the H$\beta$ flux as in Parker et al. (2011), yield a PN ionised mass of $\sim$ 0.23 $M_\odot$ (see Pottasch 1996). The relatively low CS luminosity confirms that it is an evolved CSPN, in agreement with the evolved PN nature, and consistent with the fast evolution of a high mass progenitor (see Villaver et al. 2003).  

The apparent absence of a binary companion is controversial for a bipolar nebula (De Marco 2009) as the one examined here and this study indicates that 
extreme shapes may be produced even without any significant contribution from a companion at least in the case of massive stars. 

The integrity of our adopted method for determining the CSPN luminosity and effective temperature rests on the assumptions that the nebula is optically thick and 
the nebular internal extinction is negligible.

\par
%

\section{Conclusions}

We have used our deep HST photometric data to constrain the physical parameters of the Galactic open cluster  AL~1 using the deepest CMD ever constructed 
for this cluster. Our new results, presented in Table 1, are in a close 
agreement with those from the best previous studies but with tighter errors. We confirm 
that this intermediate-age cluster is indeed one of the most distant known in our 
Galaxy. 
Furthermore, the HST data have allowed us to identify for the first time the CSPN of PHR~1315-6555, a unique PN 
proven to be a member of OC AL~1 (Parker et al. 2011). Our 
analysis indicates that, as might be expected, it is a hot blue star close to the nebular apparent centre. Our findings are of great interest since they uniquely 
provide direct measurements of the physical parameters of a Galactic CSPN in an OC with a precisely known distance. 

\section*{Acknowledgements}
The first author thanks the University of Hong Kong for the provisions
of a PhD scholarship. The second author acknowledges the support of
GRF grants 17326116 and 17300417 from the Research Grants Council of
Hong Kong that helped make this research possible. We would like to enthusiastically thank Dr. David Frew for his early discussions and input to formulating the original HST proposal in 2012.








%

%


\bsp	
\label{lastpage}
\end{document}